# Growth and Allocation of Resources in Economics:
# The Agent-Based Approach

Enrico Scalas[a,b], Mauro Gallegati[c], Eric Guerci[d], David Mas[e], and Alessandra Tedeschi[f]

[a]Università del Piemonte Orientale, Italy
[b]INFM Genova, Italy
[c]Università Politecnica delle Marche, Italy
[d]Università di Genova, Italy
[e]Université Pantheon-Assas Paris II, France
[f]Università di Roma "La Sapienza", Italy

## Abstract

Some agent-based models for growth and allocation of resources are described. The first class considered consists of conservative models, where the number of agents and the size of resources are constant during time evolution. The second class is made up of multiplicative noise models and some of their extensions to continuous-time.

## 1. Introduction

In this paper, we present a survey on some important classes of simple agent-based models used for the simulation of either growth processes or the allocation of resources in economics. These models represent systems where there are $N$ agents that can interact. The interactions can be direct and can include both two-body and many-body terms, but they can also be indirect, through some coupling and feedback mechanism with an external "field". For instance, this is the role of the banking system in some agent-based models for firm growth. Each agent $i$ is characterized by a certain quantity $s_i$, which represents either size, or wealth or another relevant quantity. The interactions determine a variation of $s_i$ as a function of time. In principle, the evolution of the system can be described both in continuous time and in discrete time. There are various paradigms helping in making the previous informal description of agent-based models more rigorous. It is worth mentioning the so-called *Interacting Particle Systems* that include, as special cases, percolation, the Ising model, the voter model, and the contact model [1]. Muchnick and Solomon, in [2], have proposed a framework taking explicitly into account causality.

In the following, we first present very simple money-exchange games where the total number of agents and the aggregate size $S = \sum s_i$ are conserved. Depending on the specific model $S$ may have the meaning of *money* and/or *wealth*. These models have been used to discuss the random allocation of resources. Then, we pass to the simplest growth models with multiplicative noise (MN). They are the basis on which the so-called *Generalized Lotka-Volterra* (GLV) models are built (Solomon, 2000). In these models, the logarithm of the size is the sum of independent and identically distributed random variables and they are directly related to the pioneering work of Gibrat[1] [3] (see also [4] and [5]) as well as to more recent studies on the applications of continuous-time random walks in finance and economics (in [6], [7], [8], [9], [10] and [11]). Moreover, these models are useful tools in the study of stochastic aggregation. Indeed, for a given period or time, MN models give a value of the aggregate that is, in its turn, a stochastic variable and, in general, the distribution of $S$ may differ from the distribution of $s_i$.

## 2. Conservative models

We consider two instances of conservative models, the One Parameter Inequality Process (OPIP) by John Angle ([12], [13], [14]) and the Maxwell-Boltzmann (MB) process by Bennati ([15], [16]). In these models, the number of agents $N$ and the aggregate size $S$ do not change with time.

The OPIP can be described as follows. Suppose that there are N players in a room, each of them with an initial amount of money, $s_i(0)$. Two players are selected by chance to play against each other. They flip a coin and the winner gets from the loser a fixed *fraction*, $0 < \omega < 1$, of the loser's money. Then, the game is iterated. If, $j$ and $k$ are the selected players at step $t$, their money at step $t+1$ is given by:

$$s_j(t+1) = s_j(t) + \omega d_{t+1} s_k(t) - \omega(1 - d_{t+1}) s_j(t)$$
$$s_k(t+1) = s_k(t) - \omega d_{t+1} s_k(t) + \omega(1 - d_{t+1}) s_j(t)$$
(1a,b)

where, $d_t$ is a Bernoulli random variable assuming the value 1 with probability ½ or the value 0 also with probability ½.

For sufficiently small values of $\omega$, the stationary probability density function (*pdf*) of size is the gamma probability density function:

---
[1] The title of the paper by Gibrat is already a *manifesto*, claiming that the law of proportional effects is able to explain inequalities in many different economical phenomena.

$$p(s) = \frac{\lambda^\alpha}{\Gamma(\alpha)} s^{\alpha-1} \exp(-\lambda s) \tag{2}$$

where, the shape parameter, $\alpha$, is approximately given by

$$\alpha \approx \frac{1-\omega}{\omega} \tag{3}$$

and the scale parameter, $\lambda$, can be obtained from the estimate of the average value of $s_i$:

$$\lambda = \frac{\alpha}{\bar{s}} \tag{4}$$

with

$$\bar{s} = \frac{1}{N}\sum_{i=1}^{N} s_i. \tag{5}$$

The MB model described in [15] and [16] is very similar to the OPIP, but there is an important difference. After the coin toss, the winner receives a *fixed* amount of money, $d$. Indebtedness is impossible. Therefore, the players who reach $s_i = 0$ cannot lose more money. If they are selected to play and they lose, they stay with no money, if they win, they get the fixed amount of money from the loser. On the contrary, in the OPIP, very poor agents always lose only a fraction of they money, and they never reach the situation $s_i = 0$. For this process, using standard tools of Markov chain theory (as described in [17], [18], [19], and [10]), it is possible to prove that the stationary distribution is given by:

$$p(s) = A\exp(-\beta s), \tag{6}$$

where:

$$\begin{aligned}\beta &= \frac{1}{\Delta s}\log\left(1 + \frac{N}{S}\Delta s\right) \\ A &= N(1-\exp(-\beta \Delta s))\end{aligned} \tag{7a,b}$$

and $\Delta s$ is the width of the histogram bins. It is interesting to remark that, if $N\Delta s/S \ll 1$, then $\beta \approx N/S$ and that this parameter plays the role of an effective inverse temperature ([20]), whereas the size $S$ is analogous to the internal energy in a gas of $N$ particles.

Recently, these models have been critically reviewed in [21]. One of the main criticisms is that they do not take into account the *free choice* of economic agents to take part in the exchange. Even so, these models are very simple and, yet, they show that in the presence of finite and fixed resources, random allocation is enough to lead to inequality. It is, perhaps, not surprising that non-specialized media emphasized results based on modifications of these models, after a conference in Kolkata [22].

## 3. Variations on the theme of proportional effects

A very simple non-conservative model that takes into account the ideas of Gibrat is the following:

$$s_i(t+1) = \eta(t)s_i(t), \tag{8}$$

where $\eta$ is a random variable always extracted from the same probability distribution ([23], [24] and [25]). In this model, there is no interaction between the agents, and the subscript $i$ can be removed from equation (8). Even in the absence of interactions, the random multiplicative model gives rise to an interesting behaviour. Let us define the log-size as $x(t) = \log(s(t))$ and the *growth rate* as $\xi = \log \eta$. Then the log-size at period $t$ is the sum of the initial log-size and of a series of independent, identically distributed random variables:

$$x(t) = x(0) + \sum_{m=0}^{t-1} \xi(n). \tag{9}$$

If the growth rate is independent from the size, the Central Limit Theorem and its generalizations apply in the large $t$ limit [26] and one gets either normal or Lévy distributed log-sizes and, therefore, lognormal or log-Lévy distributed sizes. Essentially, if the distribution of $\xi$ has a finite second moment, the limiting distribution of $s$ is lognormal otherwise it is log-Lévy. It is interesting to remark that this model gives rise to a diffusive (in the normal case) or sub-diffusive (in the Lévy regime) behaviour. Therefore there is no stationary probability density function and, starting from a situation in which all the agents are characterized by the same size, the width of the probability density function increases as a function of time. Solomon, in [27], shows that, in the normal case, for large times the probability density, $p(x,t)$, of finding a log-size $x$ at time $t$ is inversely proportional to the square root of $t$:

$$p(x,t) \sim \frac{1}{\sqrt{t}}, \tag{10}$$

and the $x$ dependence is washed out. This means that, in this regime, a power law with exponent 1 approximately gives the probability density of size:

$$p(s) \sim \frac{1}{s}. \tag{11}$$

It is now useful to consider two continuous-time extensions of equations (8) and (9). The first one is as follows. Let us replace in equation (8) 1 with a finite but small time interval, $\Delta t$, therefore we have:

$$s(t+\Delta t) - s(t) = (\eta(t) - 1)s(t). \tag{12}$$

It is possible to assume that $\eta(t)-1$ is Gaussian white noise, $\Delta W$ with constant standard deviation, $\sigma$. With the passage to the limit $\Delta t \to 0$, and transformation to log-size, equation (12) becomes:

$$dx = \sigma\, dW, \qquad (13)$$

a Langevin stochastic differential equation whose Fokker-Planck equation is the normal diffusion equation:

$$\frac{\partial p(x,t)}{\partial t} = \frac{\sigma^2}{2} \frac{\partial^2 p(x,t)}{\partial x^2}. \qquad (14)$$

The Green Function of (14) is the normal probability density function:

$$p(x,t) = \frac{1}{\sqrt{2\pi\sigma^2 t}} \exp\left(-\frac{x^2}{2\sigma^2 t}\right), \qquad (15)$$

thus leading to a log-normal distribution of size *s*.

For the second extension, let us consider a situation in which the growth shocks can arrive at random times. Equation (9) is replaced by:

$$x(t) = x(0) + \sum_{m=0}^{M(t)-1} \xi_m \qquad (16)$$

where *M(t)* is the random number of shocks (growth events) that occurred from time 0 up to time *t*. Therefore, this extension to continuous time leads to pure-jump stochastic models known as *continuous-time random walks* ([9] and [11]). The discrete-time results can be generalized to continuous time. Again, as in the discrete case, if the growth shocks are independent from the size, the Central Limit Theorem and its generalizations apply in the diffusive limit ([9], [11]) and one gets either normal or Lévy distributed log-sizes and, therefore, log-normal or log-Lévy distributed sizes. It is important to remark that, as in the discrete case, there is no statistical equilibrium probability density, as the width of *p(x,t)* continuously increases with time.

**ACKNOWLEDGEMENTS**

This work was supported by the Italian M.I.U.R. F.I.S.R. Project "High frequency dynamics of financial markets". The authors first discussed the issue of statistical equilibrium in Physics and Economics within a Thematic Institute on Complexity, Heterogeneity and Interactions in Economics and Finance sponsored by the EU EXYSTENCE Network of Excellence. E.S. wishes to acknowledge stimulating discussion with Martin Hohnisch, Sorin Solomon, Guido Germano, Thomas Lux, and John (Jack) Angle.